\documentclass[fleqn,twoside]{article}
\usepackage{espcrc2}
\usepackage{xspace}

\usepackage{graphicx}
\usepackage[figuresright]{rotating}

\newcommand{\AmS}{{\protect\the\textfont2
  A\kern-.1667em\lower.5ex\hbox{M}\kern-.125emS}}
\newcommand{\optbar}[1]{\shortstack{{\tiny(\rule[.4ex]{1em}{.1mm})}\\[-.7ex]$#1$}}
\newcommand{\BorBbar}{\kern 0.18em\optbar{\kern -0.18em B}{}\xspace}

\title{LHCb strategies for $\gamma$ from $B \to DK$ }

\author{Y. Xie
        \address[MCSD]{School of Physics, University of Edinburgh,
        Mayfield Road, Edinburgh EH9 3JZ, UK}
        (on behalf of the LHCb Collaboration)
       }

\begin{document}

\begin{abstract}
One of the most promising ways to determine the angle $\gamma$ of the CKM 
unitarity triangle is through measurement of 
the tree level processes $B \to DK$. The LHCb collaboration has
studied the potential of these decays employing the Atwood-Dunietz-Soni (ADS) and Dalitz methods, making 
use of a large sample of simulated data. For each method the expected sensitivities to the angle $\gamma$ 
are presented in this report. 

\vspace{1pc}
\end{abstract}

\maketitle

\section{INTRODUCTION}
\label{sec:1}
This report is arranged as follows. In Section~\ref{sec:1} we present a physics motivation, 
followed by a general discussion of $B \to DK$ decays 
\footnote{In this report $D$ represents a $D^0$ or  $\bar{D^0}$,
                $B$ represents a $B^{\pm}$, $B^0$ or $\bar{B^0}$
                and $K$ represents a $K^{\pm}$, $K^{*0}$ or  $\bar{K^{*0}}$.}. 
                Section~\ref{sec:2} gives a 
brief description of the LHCb detector,  the simulation and the event selection techniques.  
An introduction to the ADS method to extract the CKM 
angle $\gamma$ from $B \to DK$ decays and its application and expected performance
in LHCb is presented in Section~\ref{sec:3}. Section~\ref{sec:4} describes 
the use of the Dalitz method to extract
 $\gamma$ from $B \to DK$ decays and the expected LHCb sensitivities.   
We conclude in Section~\ref{sec:5}.

\subsection{Motivation}

In B hadron decays, 
tree level processes are generally dominated by Standard 
Model contributions while new physics mainly affects loop diagrams. 
Any difference between the CKM unitarity triangle as measured in tree 
level processes compared to loop level processes would indicate 
new physics in the flavour sector. Currently the unitarity triangle 
as determined by the tree level measurements of $\gamma$ and the CKM matrix element
$|V_{ub}|$ is consistent with the triangle determined by measurements of
the loop dominated 
parameters $\epsilon_k$, $\Delta m_d$, $\Delta m_s$ and $\sin2\beta$ \cite{Charles}. 
However, the present measurements of $\gamma$ from $B \to DK$ at B factories have 
large errors:  
$ \gamma =(92 \pm 41 \pm 11 \pm 12)^{\rm o} ({\rm BaBar})$ \cite{DalitzBabar},
$  \gamma =(53 ^{+15}_{-18} \pm 3 \pm 9 )^{\rm o} ({\rm Belle})$ \cite{DalitzBelle}.

The precision on the angle $\gamma$ from $B \to DK$ has to be improved to  
at least $5^{\rm o}$ \cite{Charles}  to match the precision of  its indirect estimate from 
a global fit to CKM parameters excluding direct measurements of $\gamma$. 

In LHCb,  the tree level decays $B^0_s \to D_s^{\pm} K^{\mp}$ can also be used to 
determine the angle $\gamma$ in a theoretically  clean way. This measurement is expected
to have a sensitivity of $14 ^{\rm o}$ with 2 ${\rm fb}^{-1}$ of data. We will 
see below that the  $B \to DK$ decays 
have a greater statistical sensitivity to the angle $\gamma$ with an
equivalent amount of data.

\subsection{Features of $B \to DK$ decays}
The CKM favoured process  $B \to \bar{D^0}K$ and disfavoured process 
$B \to D^0K$ can be described with three parameters: a weak phase
difference $\gamma \equiv arg(-(V_{ud}V^*_{ub}/(V_{cd}V^*_{cb}))$, a strong phase 
difference $\delta_B$ and the ratio of magnitudes between the disfavoured and 
favoured amplitudes,
defined as $r_B$.
If the $\bar{D^0}$ and 
$D^0$ decay to a
common final state, then the interference between the two amplitudes via
$\bar{D^0}$ and $D^0$ allows the extraction the angle $\gamma$
with several methods, as illustrated below.

It is expected that $r_B$ is small in the $B^{\pm} \to DK^{\pm}$ case
and has a value
of about 0.1 \cite{Gronau,BDecays} due to colour suppression in the 
CKM disfavoured amplitude. 
For  $ \bar{B^{0}}\to D \bar{ K^{*0}}$, both amplitudes are colour 
suppressed, therefore $r_B$ is expected to be larger
\cite{BDecays}.
 
\section{THE LHCb DETECTOR}
\label{sec:2}
The LHCb detector is a single arm spectrometer dedicated to the study of 
CP violation in B meson decays at the Large Hadron Collider, which will 
start operation at CERN in 2007. The detector 
and its expected performance is described in detail in \cite{ReopTDR}. 
Here we only emphasize that LHCb has a $94\%$ tracking 
efficiency for tracks with momentum above 10 ${\rm GeV/c}$, a $93 \%$ $K^{\pm}$
identification efficiency and a corresponding probabilty 
of $4.7\%$ for a $\pi^{\pm}$ to be misidentified as a $K^{\pm}$ for the momentum range
$2 - 100$ ${\rm GeV/c}$.

\subsection{Data simulation and event selection}
Monte Carlo simulation data produced with 
Pythia and Geant4 are used to study the trigger, the reconstruction and event selection,
which in turn allows the physics performance to be assessed.

We use event samples consisting of  260 million minimum bias event for 
trigger studies, 140 million inclusive $b\bar{b}$ events and
dedicated signal events for selection and background studies.

Sensitivities on physics parameters are obtained using fast simulation. These are
based on efficiencies, resolutions and background levels 
obtained from the full simulation.

In our studies we use the following information to discriminate between signal
 and background events: charged particle identification information based on the 
Ring Imaging Cherenkov detectors; invariant masses; 
impact parameters; transverse momenta; $\chi^2$ of decay vertices 
of the $B$, $D$, $K^*$ and $K^0_S$ particles; 
the opening angle between the momentum 
direction of a B and its flight direction; the event topology itself.

All selection cuts are optimized to reject most background events from a large 
sample of inclusive $b\bar{b}$ events while retaining a reasonable signal 
efficiency. The ratio of background to signal, $B/S$, is assessed using a separate 
inclusive   $b\bar{b}$   sample.

\section{ADS METHOD AND SENSITIVITY}
\label{sec:3}

\subsection{Description of the method }

Atwood, Dunietz and Soni (ADS)~\cite{ADS} suggested a method of determining 
$\gamma$ based on the reconstruction of  non-CP eigenstates 
for decays common to both $D^0$ and $\bar{D^0}$. An example is the hadronic final state 
$K\pi$, which  may arise from the Cabibbo favoured (CF) decay
$\bar{D^0} \to K^+ \pi^-$ or the doubly-Cabibbo suppressed (DCS) decay 
$D^0 \to K^+ \pi^-$. The relation between the CF and the DCS decay is described by a 
magnitude ratio $r^{K\pi}_D$ and a strong phase difference $\delta^{K\pi}_D$:

\begin{equation}
     r^{K\pi}_D e^{i\delta^{K\pi}_D} \equiv   
       \frac{A(\bar{D^0}\to K^- \pi^+)}{A(D^0\to K^-\pi^+)} =
       \frac{A(D^0 \to K^+ \pi^-)}{A(\bar{D^0} \to K^+ \pi^-)}. 
\end{equation} 

Therefore there are four possible $B$ decays, whose decay rates can be written as follows

$$\Gamma(B^- \to (K^- \pi^+)_D K^-) \propto $$
\begin{equation}
  1 + (r_B r^{K \pi}_D)^2 + 2r_B r^{K \pi}_D \cos(\delta_B - \delta^{K\pi}_D -\gamma), 
\end{equation}
$$\Gamma(B^- \to (K^+ \pi^-)_D K^-) \propto $$
\begin{equation}
  r_B^2 + (r^{K \pi}_D)^2 + 2r_B r^{K \pi}_D \cos(\delta_B +\delta^{K\pi}_D -\gamma), 
\end{equation}
$$\Gamma(B^+ \to (K^+ \pi^-)_D K^+) \propto $$
\begin{equation}
  1 + (r_B r^{K \pi}_D)^2 + 2r_B r^{K \pi}_D \cos(\delta_B - \delta^{K\pi}_D +\gamma), 
\end{equation}
$$\Gamma(B^+ \to (K^- \pi^+)_D K^+) \propto $$
\begin{equation}
  r_B^2 + (r^{K \pi}_D)^2 + 2r_B r^{K \pi}_D \cos(\delta_B + \delta^{K\pi}_D +\gamma), 
\end{equation}

\noindent 
where the constant of proportionality is the same in each expression. The 
relative rates of the 
four processes yield three observables which depend on five parameters $\gamma$, 
$r_B$, $r^{K\pi}_D$, $\delta_B$ and  $\delta^{K\pi}_D$, of which
$r^{K\pi}_D$ is already well known. 
It is necessary to use different $D$ decays in order to determine all parameters.

Similarly, the four-body decay $D^0\to K\pi\pi\pi$ provides three new 
observables which depend
on  $\gamma$, $r_B$, $\delta_B$ 
and two new parameters $r^{K3\pi}_D$ and  $\delta^{K3\pi}_D\,\,$\footnote{
In fact the inclusion of $D^0\to K\pi\pi\pi$ will bring many more parameters -- and
additional information -- because of the resonant substructure of the decay.
This complication is neglected in the present discussion.}.
Further information can be added by including $D$ decays to 
CP-eigenstates, such as $K^+K^-$ 
and $\pi^+\pi^-$. Each provides one more observable without introducing any
new parameters: 

\begin{equation}
\Gamma(B^- \to (h^+ h^-)_D K^-) \propto 
  1 + (r_B)^2 + 2r_B \cos(\delta_B - \gamma), 
\end{equation}

\begin{equation}
\Gamma(B^+ \to (h^+ h^-)_D K^+) \propto 
  1 + (r_B )^2 + 2r_B \cos(\delta_B  +\gamma). 
\end{equation}

\subsection{Performance with charged $B$ mesons}
Results from the B factories favour a small value of $r_B$ for charged $B$ decays~\cite{BabarRB,BelleRB}.
We set $r_B = 0.077$ for this study. The following assumptions are made: $r^{K\pi}_D = r^{K3\pi}_D = 
0.06$,  $-25^{\rm o} < \delta^{K\pi}_D < 25^{\rm o}$ and $-180^{\rm o} < \delta^{K3\pi}_D < 180^{\rm o}$.

Tab.~\ref{table:1} shows the expected signal and background yields in 2 
${\rm fb}^{-1}$ of data.
Of the 17.7 $k$  background events in the favoured $B \to (K\pi)_D K$ modes, 17.0 $k$ are from 
the decay $B \to D\pi$, which has a 13 times larger branching ratio,
with a $\pi$ misidentified as a $K$, and 0.7 $k$ are
combinatorial background events.  In the suppressed $B \to (K\pi)_D K$ modes, 
the combinatorial background dominates.

\begin{table*}[htb]
\caption{Expected signal yields $S$, number of background events $B$ 
and the rato $B/S$ in 
2 ${\rm fb}^{-1}$ of data 
for ADS decay modes of $B^{\pm}$ corresponding to 
$\delta_B=130^{\rm o}$ and $\delta^{K\pi}_D=0^{\rm o}$.
The uncertainty on the background estimates is around 60\% for
the rarest modes. }
\label{table:1}

\newcommand{\m}{\hphantom{$-$}}
\newcommand{\cc}[1]{\multicolumn{1}{c}{#1}}
\renewcommand{\tabcolsep}{2pc} 
\renewcommand{\arraystretch}{1.2} 

\begin{tabular}{@{}llll}
\hline
 decay mode                                         &     $S$     &  $B$     & $B/S$     \\
\hline
 $B^+ \to (K^+ \pi^-)_D K^+$                        &     28 $k$    &  17.7 $k$   & 0.6       \\
 $B^- \to (K^- \pi^+)_D K^-$                        &     28 $k$    &  17.7 $k$   & 0.6       \\
 $B^+ \to (K^- \pi^+)_D K^+$                        &     530     &  770      & 1.5       \\
 $B^- \to (K^+ \pi^-)_D K^-$                        &     180     &  770      & 4.3       \\
\hline
 $B^+ \to (K^+K^-/\pi^+\pi^-)_D K^+$                &     4.3 $k$   &  4.3 $k$   & 1.0       \\
 $B^- \to (K^+K^-/\pi^+\pi^-)_D K^-$                &     3.3 $k$   &  3.3 $k$   & 1.0       \\
\hline
     
\end{tabular}\\[2pt]
\end{table*}
 
Our simulation study indicates that the signal yields and background level are very 
similar 
for the $D \to K3\pi$ modes. Therefore we use the same yields and $B/S$
as in  $D \to K\pi$.

A large number of fast samples are generated based on the 
yields and background levels given above to estimate the 
statistical precision of $\gamma$. As shown in Tab.~\ref{table:2}
a precision of $5^{\rm o} - 15^{\rm o}$ for $\gamma$ is achievable 
for 2 ${\rm fb}^{-1}$ of data, depending on the parameter values 
of $\delta^{K\pi}_D$ and $\delta^{K3\pi}_D$ and with $r_B=0.077$.
Better precision can be achieved for  larger $r_B$ values.

\begin{table*}[htb]
\caption{The statistical error of $\gamma$ for different  values of $\delta^{K\pi}_D$ and 
$\delta^{K3\pi}_D$ for $r_B=0.077$.  Numbers with $^*$ are RMS values quoted for non-Gaussian 
 distribution of fit results due to close lying ambiguous solutions.
 These will disappear as the signal yields increase. 
}
\label{table:2}
                                                                                                                                                         
\newcommand{\m}{\hphantom{$-$}}
\newcommand{\cc}[1]{\multicolumn{1}{c}{#1}}
\renewcommand{\arraystretch}{1.2} 
                                                                                                                                                         
\begin{tabular}{@{}llllllll}
\hline
 $\delta^{K\pi}_D$ & $-25^{\rm o}$ & $-16.6^{\rm o}$ & $-8.3^{\rm o}$ & $0^{\rm o}$ & 
 $8.3^{\rm o}$ & $16.6^{\rm o}$ & $25^{\rm o}$ \\
\hline
$\delta^{K3\pi}_D=-180^{\rm o}$ &  $8.6^{\rm o}$ &  $7.5^{\rm o}$ &  $6.5^{\rm o}$ &
 $6.8^{\rm o*}$ &  $7.2^{\rm o*}$ &  $7.3^{\rm o*}$ &  $6.0^{\rm o*}$ \\
$\delta^{K3\pi}_D=-120^{\rm o}$ &  $6.0^{\rm o}$ &  $6.3^{\rm o}$ &  $6.3^{\rm o}$ &
 $6.4^{\rm o}$ &  $6.2^{\rm o}$ &  $6.2^{\rm o}$ &  $4.7^{\rm o}$ \\
$\delta^{K3\pi}_D=-60^{\rm o}$ &  $8.0^{\rm o}$ &  $7.9^{\rm o}$ &  $8.1^{\rm o}$ &
 $7.8^{\rm o}$ &  $7.4^{\rm o}$ &  $6.7^{\rm o}$ &  $6.2^{\rm o}$ \\
$\delta^{K3\pi}_D=0^{\rm o}$ &  $10.3^{\rm o*}$ &  $11.1^{\rm o*}$ &  
$12.^{\rm o*}$ & $11.5^{\rm o*}$ &  $12.1^{\rm o*}$ &  $13.1^{\rm o*}$ &  $13.0^{\rm o*}$ \\
$\delta^{K3\pi}_D=60^{\rm o}$ &  $9.1^{\rm o}$ &  $10.6^{\rm o}$ &  $11.2^{\rm o}$ &
 $12.9^{\rm o}$ &  $13.4^{\rm o*}$ &  $15.0^{\rm o*}$ &  $15.2^{\rm o*}$ \\
$\delta^{K3\pi}_D=120^{\rm o}$ &  $11.6^{\rm o*}$ &  $11.3^{\rm o*}$ &  $11.8^{\rm o*}$ &
 $11.0^{\rm o*}$ &  $10.9^{\rm o*}$ &  $11.1^{\rm o}$ &  $10.9^{\rm o}$ \\
$\delta^{K3\pi}_D=180^{\rm o}$ &  $8.5^{\rm o}$ &  $7.4^{\rm o}$ &  $6.5^{\rm o}$ &
 $6.8^{\rm o}$ &  $7.1^{\rm o}$ &  $7.3^{\rm o}$ &  $6.5^{\rm o}$ \\
\hline
\end{tabular}\\ 
\end{table*}

LHCb is also investigating the feasibility to include the decay
$B^{\pm} \to D^{*0} K^{\pm}$ in the ADS analysis. An attractive 
feature of the $D^{*0}$ meson is that it can decay to two final
states $D^0\pi^0$ and $D^0\gamma$. These have opposite 
CP eigenvalues, which lead to a difference of $\pi$ in their
strong phases.
This is a useful constraint if the two decays can be 
distinguished experimentally. However, these decays are difficult to
fully reconstruct in LHCb because the detection efficiency of a soft photon in
the electromagnetic calorimeter is very low while the background is enormous.
Alternative approaches which employ a 
partial reconstruction or which make use of the event
topology to reconstruct the momentum of the $\pi^0/\gamma$ are 
under study.

\subsection{Performance with neutral $B$ mesons}
The same method can also be applied to the decay
 $ \bar{B^{0}}\to D \bar{ K^{*0}}$
with $D \to K\pi$, $KK$ or $\pi\pi$. Assuming $r_B =0.4$, $55^{\rm o}<\gamma < 105^{\rm o}$
and $-20^{\rm o}<\delta_B<20^{\rm o}$, the expected signal yields in 2 ${\rm fb}^{-1}$ 
of data and the background-to-signal ratio $B/S$ are given in 
Tab.~\ref{table:3}.
The corresponding statistical precision of the angle $\gamma $ is 
expected to be $7^{\rm o} - 10^{\rm o}$.  

\begin{table*}[htb]
\caption{Expected signal yields in 2 ${\rm fb}^{-1}$ of data and  
the background-to-signal ratio $B/S$ for the ADS modes of neutral $B$ mesons. 
Upper limits with $90\%$ confidence level are quoted for $B/S$. }
\label{table:3}

\newcommand{\m}{\hphantom{$-$}}
\newcommand{\cc}[1]{\multicolumn{1}{c}{#1}}
\renewcommand{\tabcolsep}{2pc} 
\renewcommand{\arraystretch}{1.2} 

\begin{tabular}{@{}lll}
\hline
 decay mode                                  &     $S$     & $B/S$     \\
\hline
 $B^0 \to (K^- \pi^+)_D K^{*0} + c.c.$       &     3400    & $<$0.3 \\       
 $B^0 \to (K^+ \pi^-)_D K^{*0} + c.c. $      &     500    & $<$1.7 \\
 $B^0 \to (K^+K^- / \pi^+ \pi^-)_D K^{*0} + c.c.$       &     500    & $<$1.4 \\
\hline
\end{tabular}\\
\end{table*}

\section{DALITZ METHOD AND SENSITIVITY}
\label{sec:4}

\subsection{Three-body $D$ decay}
This method for determining $\gamma$ was proposed in  \cite{DalitzGiri}. 
It makes use of the decay $B^{\pm}\to DK^{\pm}$ followed
by a multibody $D$ decay into a CP eigenstate
Here we explain the basic idea using $D\to K^0_S \pi^+ \pi^-$ as an example.
The Dalitz phase space of this decay $D\to K^0_S \pi^+ \pi^-$  
can be fully parameterized with two effective masses 
$m^2_+ \equiv m^2(K^0_S \pi^+)$ and $m^2_- \equiv m^2(K^0_S \pi^-)$. 
The $D^0$ and $\bar{D^0}$ decay amplitudes can be written as 
functions $f(m^2_- , m^2_+ )$ and $f(m^2_+ , m^2_- )$.

The total Dalitz decay amplitudes, defined as  $A^-\equiv A(B^-\to 
(K^0_S\pi^+\pi^-)_D K^-)$
and $A^+\equiv A(B^+\to (K^0_S\pi^+\pi^-)_DK^+)$,
 are sums of contributions via $D^0$ and $\bar{D^0}$:

\begin{equation}
A^-= f(m^2_- , m^2_+ ) + r_B e^{i(-\gamma + \delta_B)}f(m^2_+ , m^2_- ),
\end{equation}

\begin{equation}
A^+= f(m^2_+ , m^2_- ) + r_B e^{i(\gamma + \delta_B)}f(m^2_- , m^2_+ ).
\end{equation}

In the isobar model \cite{isobarCLEO}  $f(m^2_+ , m^2_- )$ is a coherent sum of contributions 
of different resonances:

\begin{equation}
f(m^2_+ , m^2_- ) = \sum_{j=1}^N a_je^{i\alpha_j}A_j (m^2_+ , m^2_- )+ be^{i\beta},
\end{equation}

\noindent 
where $a_j$, $\alpha_j$, $b$ and $\beta$ are model parameters which have been measured 
well at the B factories  \cite{DalitzBabar,DalitzBelle}. The $B^{\pm}$ decay rates are given by 

$$\Gamma^-(m^2_+ , m^2_- ) = |f(m^2_- , m^2_+ )|^2 + r_B^2 |f(m^2_+ , m^2_- )|^2$$
\begin{equation}
+ 2r_BRe[f^*(m^2_- , m^2_+ ) f(m^2_+ , m^2_- )e^{i(-\gamma+\delta_B)}],
\end{equation}
$$\Gamma^+(m^2_+ , m^2_- ) = |f(m^2_+ , m^2_- )|^2 + r_B^2 |f(m^2_- , m^2_+ )|^2$$
\begin{equation}
+ 2r_BRe[f^*(m^2_+ , m^2_- ) f(m^2_- , m^2_+ )e^{i(\gamma+\delta_B)}].
\end{equation}

We can see that the interference terms are sensitive to $\gamma$, which can be determined
by measuring $\Gamma^-(m^2_+ , m^2_- )$ and $\Gamma^+(m^2_+ , m^2_- )$ across 
the Dalitz phase space.

To reconstruct the $B^{\pm}\to (K^0_S\pi^+\pi^-)_D K^{\pm}$ events is challenging with 
the LHCb detector as only $25\%$ of the $K^0_S$ particles decay inside the active region of
the vertex detector. 
The expected  signal yield in 2 ${\rm fb}^{-1}$ of data will
vary between 1.5 $k$  and 5 $k$, depending on how many of the $K^0_S$ decays 
successfully found offline can be reconstructed within the CPU constraints of the 
High Level Trigger.
A full simulation has been performed to estimate 
the background level. The combinatorial background is expected to 
contribute less than 3.5 $k$ events and contamination from $B^{\pm}\to (K^0_S\pi^+\pi^-)_D \pi^{\pm}$ 
is expected to be around 1200 events.
Our present sensitivity studies for $\gamma$ do not include background, and
do not take into account the non-flat 
acceptance efficiency in the Dalitz space, which is expected as a result of 
the trigger and offline selection. Under these assumptions, 
a statistical precision of $\sigma_ {\gamma} \approx 8^{\rm o} - 16^{\rm o}$ 
is achievable in  2 ${\rm fb}^{-1}$ of data. The actual statistical 
precision will depend on the final background level and on $r_B$.

LHCb is also investigating the decays $B^{\pm}\to (K^0_S K^+ K^-)_D K^{\pm}$ and 
$B^0\to (K^0_S\pi^+\pi^-)_D K^{*0}$.  

\subsection{Four-body $D$ decay}
The Dalitz method can be extended from three to four-body $D$ decays.
 In this case five parameters are required to describe the Dalitz phase space. 
The $D$ decay model has been studied in the FOCUS experiment \cite{FourBodyFocus}
and our $\gamma$ sensitivity studies are based on their results.

Assuming a branching ratio of 
 $B(B^{\pm}\to (K^+ K^- \pi^+ \pi^-)_D K^{\pm}) =9.5 \times 10 ^{-7}$, 
our full simulation yields 1.7 $k$ events in 2 ${\rm fb}^{-1}$ of data. 
Based on this we estimate a 
statistical precision for $\gamma$ to be $\sigma_{\gamma} \approx 14^{\rm o}$,
where we have assumed $\gamma = 60^{\rm o}$, $r_B = 0.08$ and $\delta_B=130^{\rm o}$, 
 and have not yet included 
background and detector effects. Results from more recent studies 
can be found in  \cite{JonathFourBody}. 

We are also studying an amplitude analysis for  
$B^{\pm}\to (K^\pm \pi^\mp \pi^+ \pi^-)_D K^{\pm}$. Compared with the ADS
analysis discussed in Section~\ref{sec:3}, the advantage of this
method is that it takes into account the variation of the strong phase 
$\delta_D^{K3\pi}$ in the Dalitz phase space.

\subsection{Systematic errors}
The biggest systematic uncertainty in the Dalitz method arise from the model dependence of
the  $D$ decay. In the present B-factory 
$B^{\pm}\to (K^0_S  \pi^+ \pi^-)_D K^{\pm}$ analyses this uncertainty is
around $10^{\rm o}$. This error is expected to reduce significantly through
exploitation of the coherently produced $D$ mesons available at CLEO-c~\cite{CLEOc}
and BES~\cite{BES}. A discussion of how these data may be used in a model independent
analysis can be found in~\cite{DalitzGiri,Bondar}.   Similar techniques
can be used for the four-body decay mode, where LHCb also expects large numbers
of flavour-tagged $D$ decays for use in model calibration.

\section{CONCLUSIONS}
\label{sec:5}
We have shown that LHCb will be able to extract the CKM angle $\gamma$ in several ways
with $B\to DK$ decays.
The combined result is expected to have a precision of around 
$5^{\rm o}$  with 2~${\rm fb}^{-1}$ of data.
Such a result will make it possible to compare the LHCb measurement of 
the angle $\gamma$ with the indirect determination from a CKM fit and thereby perform
a stringent test of the Standard Model.   Together with improvement on the $|V_{ub}|$ 
measurement at the B factories this will provide a precise reference Unitarity 
Triangle against which new physics searches can be compared.

\end{document}